\documentclass[12pt,preprint]{aastex}

\shorttitle{}
\shortauthors{}

\begin{document}

\title{The double AGN in NGC 6240 revealed through 3-5~$\mu$m spectroscopy\altaffilmark{*}}

\altaffiltext{*}{Based on observations collected at the European Southern Observatory, 
Chile (proposal 73.B-0574)}

\author{
G. Risaliti\altaffilmark{1,2}, E. Sani\altaffilmark{3},
R. Maiolino\altaffilmark{1}, A. Marconi \altaffilmark{1},
S.~Berta\altaffilmark{4},
V.~Braito\altaffilmark{5},
R.~Della~Ceca\altaffilmark{5},
A.~Franceschini\altaffilmark{4},
M.~Salvati\altaffilmark{1}
} 
\email{grisaliti@cfa.harvard.edu}

\altaffiltext{1}{INAF - Osservatorio di Arcetri, L.go E. Fermi 5,
Firenze, Italy}
\altaffiltext{2}{Harvard-Smithsonian Center for Astrophysics, 60 Garden St. 
Cambridge, MA 02138 USA}
\altaffiltext{3}{Dipartimento di Astronomia, Universit\`a di Firenze,
  L.go E. Fermi 2, I-50125 Firenze, Italy
}
\altaffiltext{4}{Dipartimento di Astronomia, Universit\`a di Padova, Italy
}
\altaffiltext{5}{
INAF - Osservatorio Astronomico di Brera, Milano, Italy
}
\begin{abstract}
We present 3-5~$\mu$m spectroscopy of the interacting system NGC~6240,
showing the presence of two active galactic nuclei.
The brightest (southern) nucleus shows up with a starburst-like
emission, with a prominent 3.3~$\mu$m emission feature. However, the
presence of an AGN is revealed by the detection of a  broad Br$\alpha$ emission line, with a width of $\sim1,800$~km~s$^{-1}$.
The spectrum of the faintest (northern) nucleus shows typical AGN features,
such as a steep continuum and broad absorption features in the M-band.

We discuss the physical properties of the dusty absorbers/emitters, 
and show that in both nuclei the AGN is dominant in the 3-5~$\mu$m band,
but its contribution to the
total luminosity is small (a few percent of the starburst emission).
\end{abstract}


\section{Introduction}

L-band spectroscopy ($\sim3-4~\mu$m) of Ultraluminous Infrared Galaxies (ULIRGs)
is a powerful tool to disentangle the starburst and AGN contributions
to the huge ($>10^{12}~L_\odot$) infrared luminosity. 
Several spectral features can be used as indicators of one of the two components
(e.g. Imanishi \& Dudley~2000, Risaliti et al.~2005, hereafter R05). More specifically:\\
- a large equivalent width of the 3.3~$\mu$m PAH emission feature ($EW_{3.3}\sim$100~nm)
is typical of starburst-dominated sources;\\
- a strong absorption feature at 3.4~$\mu$m ($\tau_{3.4}>0.2$), due to
alyphatic hydrocarbon grains, is an
indicator of an obscured AGN;\\
- a steep red continuum ($f_\lambda \propto \lambda^\Gamma, \Gamma>2$)
suggests the presence of an obscured, reddened AGN.

M-band spectra ($4.5-5~\mu$m) of ULIRGs are available only for a small number of sources.
From the analysis of the M-band emission of the nearby obscured AGN NGC~4945
(Spoon et al.~2003) strong absorption features due to CO ices are expected for
obscured AGNs.

NGC~6240 is an interacting system consisting of two nuclei with a separation of 1.8~arcsec 
(Fried \& Schulz~1984), corresponding to 
$\sim$800~pc\footnote{We adopt $H_0=70$~km~s$^{-1}$~Mpc$^{-1}$, e.g Spergel et al.~2003).}
and with a total infrared luminosity $L_{IR}=10^{11.8}~L_\odot$
(Genzel et al.~1998). It is optically classified as a LINER (Rafanelli et al.~1997), 
and no indications of
an AGN, such as broad Pa$\alpha$ or Br$\gamma$ lines, 
are present in the near-IR. L-band spectroscopy, performed with a four meter
class telescope, did not resolve the two nuclei, and 
showed a typical starburst emission, with a flat continuum,
a strong 3.3~$\mu$m emission feature (EW$\sim$70~nm), 
and no absorption features (Imanishi \& Dudley~2000).
In the hard X-rays  
the AGN emission is dominant above 10~keV (Vignati et al.~1999), while at lower
energy only the reflected component is visible, due to the high column density 
($N_H\sim2\times10^{24}$~cm$^{-2}$) obscuring the direct component.
In a recent {\em Chandra} observation the two nuclei are clearly separated,
and both show a prominent iron $K\alpha$ emission line, with $EW>1$~keV, indicating
the presence of an AGN in both nuclei (Komossa et al.~2003). This is the first 
clear detection of a double AGN in an interacting system.

Here we present new VLT L-band and M-band spectra of the two nuclei of NGC~6240,
both showing clear AGN features.  
\section{Reduction and Data Analysis}
NGC~6240 was observed with the Infrared Spectrometer And Array Camera (ISAAC)
at the Antu Unit (UT1) of the Very Large Telescope on Cerro Paranal, Chile,
on July 31, 2005, with photometric conditions and a seeing of 0.6~arcsec
in the infrared. 
We performed the observations in low resolution mode, with the L-band
(2.9-4.2~$\mu$m) and M-band (4.2-5.1~$\mu$m) filters, with a 0.6x20 arcsec slit, 
oriented in order to  obtain spectra of both nuclei.
The spectral resolution was $\lambda/\Delta\lambda\sim600$ in the L-band
and  $\lambda/\Delta\lambda\sim800$ in the M-band. The on-source observation times
are 45~min in the L-band and 60~min in the M-band.

In order to avoid saturation due to the high background ($\sim3.9$~mag per arcsec$^2$ in
L-band; $\sim1.2$~mag in the M-band) 
the spectra were taken in chopping mode, with single exposures
of 0.56~s.
The spectra were then aligned and merged into a single image. We performed
a standard data reduction, consisting of flat-fielding, background subtraction,
and spectrum extraction, using the IRAF~2.11 package.
In order to facilitate the background subtraction,
the observations were performed by nodding the source along the slit (with
a throw of 20 arcsec)
Both nuclei are unresolved, but the tail of their emissions
overlap. In particular, the contribution of the brightest nucleus to the emission
in the extraction region of the faint nucleus is not negligible. In order
to correct for this contamination, we subtracted from the faint nucleus a
background extracted from the region symmetric to the extraction region with respect to
the emission peak of the bright nucleus (Fig.~1). 
\begin{figure}
\includegraphics[width=8.5cm]{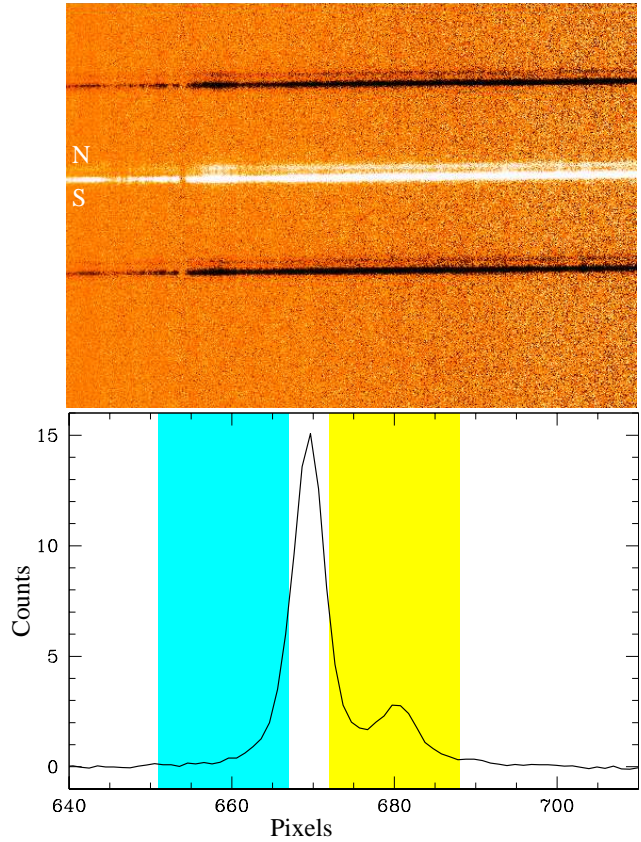}
\caption{Upper panel: L-band spectrum of NGC~6240. The 2 nuclei are clearly separated.
Lower panel: Integrated emission along the slit. We extracted the spectrum of the
faintest nucleus from the cyan shaded region on the right.
The background was extracted from the yellow shaded region on the left,
symmetric with the source region with respect to the emission peak of the
brightest nucleus.}
\end{figure}

We acquired the spectra of the standard star (HR~5249, spectral type B2V)
immediately after the target observations. 
Corrections for sky absorption and instrumental response were obtained
from the spectra of the standard star, divided by its intrinsic emission,
assumed to be a pure Raleigh-Jeans spectrum.

In order to obtain a precise absolute calibration, we took into account
aperture effects by analyzing the profiles of both the targets and the
standard stars along the slit. We assumed a Gaussian profile and we estimated
the fraction of flux inside the slit assuming a perfect centering.
We estimate this procedure to have an error $<10$\% for
the L-band and $<20$\% 
for the M-band. The precision of the absolute calibration is confirmed
by (a) the agreement within 10\% with the L-band spectrum of Imanishi \& Dudley (2000), 
and (b) the cross calibration between our L-band and M-band spectra, which match
within 15\%.

The atmospheric transmission is not constant in the 3-5~$\mu$m range.
In several narrow spectral intervals in the M-band, and one in the L-band, the 
transparency is too low to obtain significant data. We excluded these
intervals from our spectral analysis.
In other wavelength ranges, such as between 2.9 and 3.2~$\mu$m, and in the M-band
in general, the atmospheric conditions vary in short timescales. This implies
that the sky features are not completely removed with the division by the calibration star.
We rebinned the spectral channels in these intervals in order to have 
significant spectral points. 
Finally, for the spectral intervals with good atmospheric transmission
and high signal-to-noise (such as the brighest nucleus in the 3.4-4.1~$\mu$m interval) 
we chose the rebinning in order to have a 
spectral element covering a wavelength interval of $\sim30$~\AA,
a factor 2 smaller than the spectral resolution.

We fitted the spectra  with a 
power law continuum and broad Gaussian absorption and emission lines. 
We left the relative normalizations between the L and M bands free to vary 
within the uncertainties given above.
%
The results are shown in Fig.~2.

\begin{figure*}
\includegraphics[width=15cm]{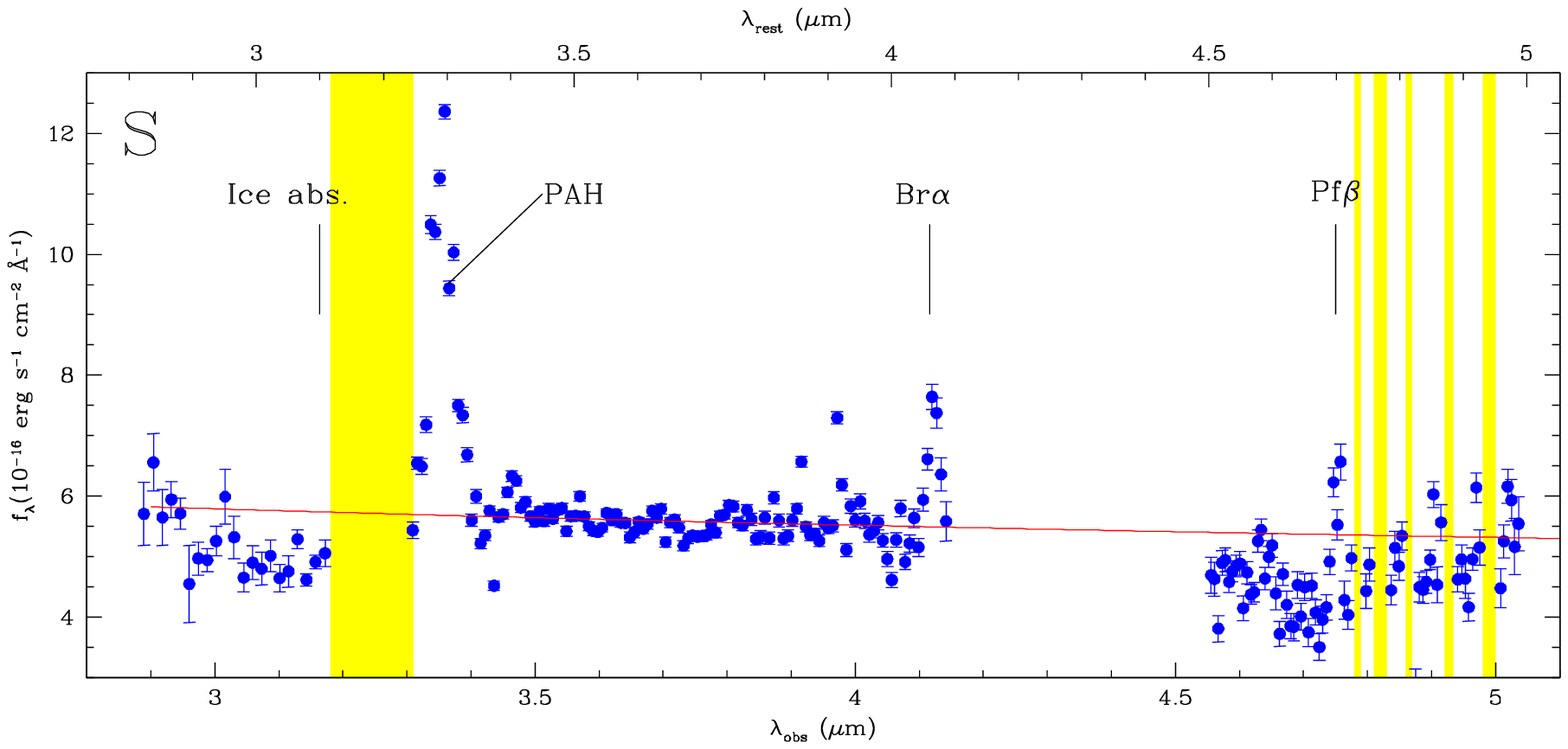}
\includegraphics[width=15cm]{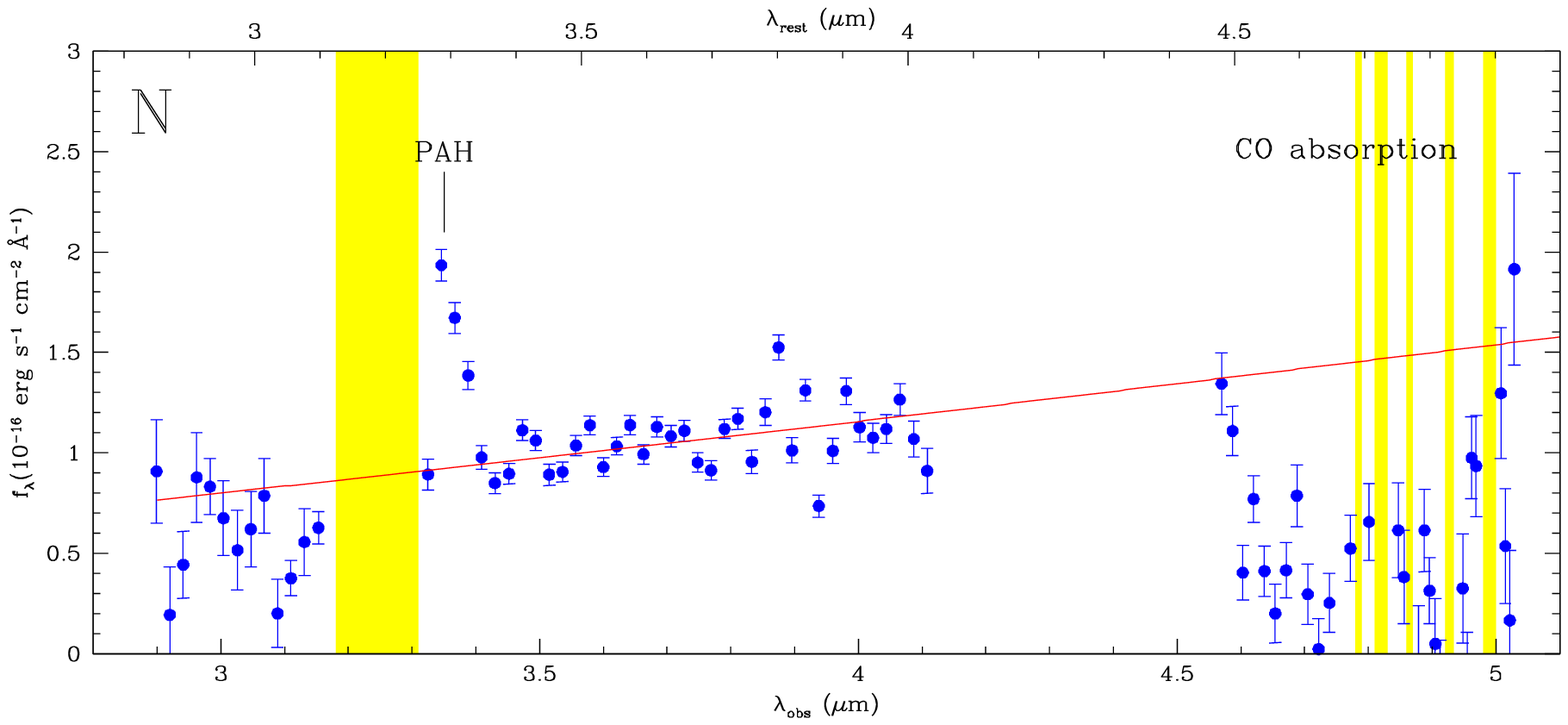}
\caption{Complete L-band and M-band spectra of the two nuclei of NGC~6240. 
We rescaled the M-band spectra
in order to match the normalization of
the L-band spectrum (see text for details). 
The yellow shaded regions show the wavelength intervals with bad atmospheric transmission.
The main spectral features are labeled.
}
\end{figure*}

\subsection{NGC~6240S}
The 3-5~$\mu$m spectral properties of the brightest, southern nucleus are the following:\\
$\bullet$ A flat $\lambda-f_\lambda$ continuum, with a slope ($\Gamma(S)=-0.3\pm0.1$)
typical of both starbursts and unobscured AGNs (R05).\\
$\bullet$ An emission feature at 3.3~$\mu$m, with $EW_{3.3}(S)=49\pm5$~nm, a factor $\sim2$
smaller than in pure starbursts (Imanishi \& Dudley~2000, R05), thus indicating the presence 
of an AGN.\\
$\bullet$ A broad Br$\alpha$ line at 4.05~$\mu$m (rest frame) with a full width at
half maximum corresponding to a velocity of 1,800$\pm200$~km~s$^{-1}$. 
This is a clear, direct evidence of the presence of an AGN in this nucleus.
The Pf$\beta$
line at 4.65~$\mu$m is also detected, but the possible atmospheric turbolence and 
CO absorption bands prevent us from analyzing the main line properties, such as profile and
equivalent width.\\
$\bullet$ A 3.1~$\mu$m ice absorption feature, with $\tau_{3.1}(S)=0.2\pm0.05$, and
no clear absorption features at 3.4~$\mu$m.\\
$\bullet$ Possible absorption features due to CO ice in the 4.5-4.8~$\mu$m range.
The strength of these features depends on the assumed continuum level.
Given the concave shape of the spectrum, and that no strong emission feature
is known around 4.6~$\mu$m, we chose the normalization of the M-band spectrum
in order to obtain a good match between the extrapolation of the L-band continuum and
the highest points in the 4.5-4.6~$\mu$m interval.
The above analysis clearly indicates the presence of both an AGN and a starburst as
sources of the 3-5~$\mu$m emission. 
\subsection{NGC~6240N}
The fainter, northern nucleus also shows a 3-5~$\mu$m emission with clear contributions 
from both an AGN and a starburst:\\
$\bullet$ The continuum is rather steep ($\Gamma(N)=1.5\pm0.1$). Such a steepness is not
typical either of starburst-dominated sources nor of unobscured AGNs, and is 
observed only in objects hosting a heavily obscured, reddened AGN (R05).\\
$\bullet$ The PAH emission feature at 3.3~$\mu$m is present, with $EW(N)=48\pm3$~nm,
a factor of 2 smaller than in starburst-dominated sources,
clearly indicating the presence of a starburst, diluted by an AGN component.\\
$\bullet$ The M-band spectrum can be interpreted as dominated by deep absorption features due
to CO ice. Again, this interpretation is not unique, due to the uncertainty 
on the continuum level. However, if the continuum is assumed to be lower (so
decreasing the strength of the absorption features) a strong emission feature at 4.6~$\mu$m would
emerge, with no obvious explanation.
\section{Discussion}
We have shown that 
both AGN and starburst components are required to explain the 3-5 micron 
emission from both nuclei of NGC 6240.
Here we quantitatively estimate the relative
contribution of the two nuclei to the L-band and bolometric luminosities of the
source, and we discuss the physical conditions of the dusty absorber/emitter.

{\bf 1. Relative AGN/starburst contributions}. In order to estimate the relative
contribution of the AGN to the 3-5$\mu$m emission in the two nuclei, we use
a simple model to disentangle the two components, as shown in R05.
We assume that the intrinsic spectrum of the AGN, $f_\lambda$(AGN), is a power law
with a spectral index $\Gamma=-0.5$, in agreement with the L-band spectra of
ULIRGs dominated by unabsorbed AGNs (R05). 
The observed AGN spectrum is obtained from the intrinsic spectrum, absorbed by
a wavelength-dependent optical depth,
$\tau(\lambda)\propto\lambda^{-1.75}$ (Cardelli et al.~1989).
The starburst component $f_\lambda$(SB), is modelled by a continuum with $\Gamma=-0.5$ (as for
unobscured AGNs) plus a broad emission feature at 3.3~$\mu$m having EW$_{3.3}$=100~nm.
This is in agreement with the average observed L-band spectra of starburst-dominated ULIRGs
(Imanishi \& Dudley~2000, R05).
The ULIRG emission, $f_\lambda$, is obtained as the combination of the two components:

\begin{equation}
f_\lambda=\alpha f_\lambda({\rm AGN}) e^{-\tau(\lambda)} + (1-\alpha) f_\lambda({\rm SB})
\end{equation}

where $\alpha$ is the fraction of the {\it intrinsic} L-band luminosity due to the AGN.
We express the optical depth as 
$\tau(\lambda)=\tau_L\times (\lambda/3.5\mu$m$)^{-1.75}$, where $\tau_L$ is
the optical depth at 3.5~$\mu$m. Therefore, the model has two
free parameters, $\alpha$ and $\tau_L$.
From the composite spectrum, the equivalent width of the 3.3~$\mu$m feature and the
continuum slope can be computed as a function of the above parameters.
Therefore, we can find the unique solution for $\tau_L$ and $\alpha$
which reproduces the observed values of $EW_{3.3}$ and $\Gamma$.
We obtain $\tau_L(S)=0.6
\pm0.2$, $\alpha(S)=0.6\pm0.1$, and
$\tau_L(N)=2.0\pm0.2$, $\alpha(N)=0.9\pm0.03$.
The quoted errors are based on the measured errors on $EW_{3.3}$ and $\Gamma$.
An additional source of systematic errors could be the uncertainty on the intrinsic AGN and
SB spectral slope, and on the intrinsic EW in SBs. If these are taken into account, our errors could
be larger by a factor $\sim$2\footnote{For further 
details on this model and on errors estimates, we refer to 
R05, where the model is also applied to a larger sample of ULIRGs.}.

In order to estimate the relative AGN/SB contribution to the bolometric luminosity,
we need to take into account the ratio, $K$, between the contribution of the
L-band emission to the bolometric luminosity in a pure AGN and that in a 
pure starburst.
From the average L-band to bolometric ratio in pure starbursts ($\sim2\times10^{-3}$,
R05) and in quasars ($\sim0.2$, from the spectral energy distribution
of Elvis et al.~1994), we obtain $K\sim100$.
The contributions of the AGN component to the bolometric luminosity,
\begin{equation} 
\alpha_{BOL}=\alpha/(\alpha+K(1-\alpha))
\end{equation}
are then $\alpha_{BOL}(S)=2\pm0.5$\% 
and $\alpha_{BOL}(N)=10\pm2$\%. 
The uncertainty on $K$ can be as large as a factor of 3-5. However, it is easily seen
from Eq.~2 that this does not affect our main conclusion, which is $\alpha_{BOL}<<1$, provided
that $K>>1$. 

Summarizing, the interesting result which emerges from the above
analysis is that a moderately absorbed AGN is present in both sources, with a dominant
contribution to the L-band emission, but a minor contribution to the bolometric luminosity.
\\
{\bf 2. Physical properties of the dusty emitter/absorber}.
The above analysis suggests a moderate L-band absorption of the AGN component in both
nuclei. Assuming a standard extinction curve ($A_V\sim25~A_L$, Cardelli, Clayton \& Mathis~1989) we
have $A_V(S)\sim$15 and $A_V(N)\sim$50. This is in agreement with the estimates of
Lutz et al.~(2003) based on the mid-IR spectrum of both nuclei (and, therefore, dominated
by the emission of the S nucleus), and 
with the absence of other broad lines in the optical and near-IR spectra,
which suffer high extinction and dilution by the starburst component. 
Instead, the Br$\alpha$ line is obscured by only $\tau(S)\sim0.6$, and lies in a spectral
region where the AGN emission is dominant. This makes Br$\alpha$ an excellent atomic emission line
to discover obscured AGN, as first noticed by Lutz et al.~(2000).

More in general, we note that in sources such as NGC~6240, where the AGN is obscured by
$A_V>$20, $N_H>10^{24}$~cm$^{-2}$, and the starburst dominates the bolometric emission,
the AGN emission is dominant in two spectral windows only: the infrared region
between 3 and 5-8~$\mu$m,
and the hard X-ray region, at E$>$10~keV. In the whole wavelength range between the L-band and
the hard X-rays the AGN is completely obscured, while in the mid and far infrared its emission
is strongly diluted by the starburst component.

The estimated values for the extinction are 
smaller by a factor about 50 than the value expected from the X-ray spectra, $A_V\sim1000$, 
if a Galactic dust-to-gas ratio is assumed 
($A_V/N_H=1.7\times10^{21}$~mag$^{-1}$~cm$^2$, Bohlin, Savage \& Drake~1978).
This discrepancy can be explained in two ways:\\
a) a lower than Galactic dust-to-gas ratio. This is commonly found in
Seyfert galaxies (Maccacaro, Elvis \& Perola~1984, Maiolino et al.~2001) and is probably common 
also among higher redshift quasars (Risaliti \& Elvis~2005).
The low ratio can be due either to the presence of a dust-free region of gas (this is
expected if part of the absorbing gas lies within the dust sublimation radius) or to
dust grains with on-average larger sizes than in the Galaxy (Maiolino, Marconi \& Oliva~2002)\\
b) The amount of absorbing gas/dust towards the X-ray emitting region could be higher than towards the
more extended L-band emitting region. The X-rays are expected to be emitted within a few tens of Schwarzschild
radii from the central black hole of the AGN, i.e. within $\sim10^{-2}$~parsec for a 10$^8$~$M_\odot$ black hole,
while the hot dust region must be at a distance $R$ larger than the sublimation 
radius (for an AGN with an intrinsic optical/UV luminosity of 10$^{46}$~erg~s$^{-1}$, $R>$1~parsec). 

\section{Conclusions}
We have presented 3-5~$\mu$m low resolution spectra of the two nuclei in the Ultraluminous
Infrared Galaxy NGC~6240, showing clear evidence of the presence of an AGN in both nuclei.
This confirms the early detection of the double AGN obtained in the hard X-rays with {\em Chandra}
(Komossa et al.~2003).

In the southern, brighter nucleus a broad Br$\alpha$ emission line is detected, the only known BLR
evidence in the spectrum of this source.

The northern nucleus shows a steep L and M band emission, typical of a reddened ($\tau_L\sim$2) AGN,
with possible strong CO absorption features in the M-band.

In both nuclei the AGN emission dominates in the 3-5~$\mu$m band, but its contribution to the bolometric
luminosity is small. The nuclear activity in sources like NGC~6240, 
i.e. where the AGN is faint (compared to the starburst)
and obscured, is non-negligible only in the infrared between 3 and $\sim5-8$~$\mu$m and in the hard X-rays.
At other wavelength the AGN is either obscured or strongly diluted by the starburst emission.
\acknowledgements



\begin{thebibliography}{}
\bibitem[Bohlin et al.(1978)]{1978ApJ...224..132B} Bohlin, R.~C., Savage, 
B.~D., \& Drake, J.~F.\ 1978, \apj, 224, 132 
\bibitem[Cardelli et al.(1989)]{1989ApJ...345..245C} Cardelli, J.~A., 
Clayton, G.~C., \& Mathis, J.~S.\ 1989, \apj, 345, 245 
\bibitem[Elvis et al.(1994)]{1994ApJS...95....1E} Elvis, M., et al.\ 1994, 
\apjs, 95, 1 
\bibitem[Fried \& Schulz(1983)]{1983A&A...118..166F} Fried, J.~W., \& 
Schulz, H.\ 1983, \aap, 118, 166 
\bibitem[Genzel et al.(1998)]{1998ApJ...498..579G} Genzel, R., et al.\ 
1998, \apj, 498, 579 
\bibitem[Imanishi \& Dudley(2000)]{2000ApJ...545..701I} Imanishi, M., \& 
Dudley, C.~C.\ 2000, \apj, 545, 701 
\bibitem[Komossa et al.(2003)]{2003ApJ...582L..15K} Komossa, S., Burwitz, 
V., Hasinger, G., Predehl, P., Kaastra, J.~S., \& Ikebe, Y.\ 2003, \apjl, 
582, L15 
\bibitem[Lutz et al.(2000)]{2000ApJ...530..733L} Lutz, D., et al.\ 2000, 
\apj, 530, 733 
\bibitem[Lutz et al.(2003)]{2003A&A...409..867L} Lutz, D., Sturm, E., 
Genzel, R., Spoon, H.~W.~W., Moorwood, A.~F.~M., Netzer, H., \& Sternberg, 
A.\ 2003, \aap, 409, 867 
\bibitem[Maccacaro et al.(1982)]{1982ApJ...257...47M} Maccacaro, T., 
Perola, G.~C., \& Elvis, M.\ 1982, \apj, 257, 47 
\bibitem[Maiolino et al.(2001)]{2001A&A...365...28M} Maiolino, R., Marconi, 
A., Salvati, M., Risaliti, G., Severgnini, P., Oliva, E., La Franca, F., \& 
Vanzi, L.\ 2001, \aap, 365, 28 
\bibitem[Maiolino et al.(2001)]{2001A&A...365...37M} Maiolino, R., Marconi, 
A., \& Oliva, E.\ 2001, \aap, 365, 37 
\bibitem[Rafanelli et al.(1997)]{1997A&A...327..901R} Rafanelli, P., 
Schulz, H., Barbieri, C., Komossa, S., Mebold, U., Baruffolo, A., \& 
Radovich, M.\ 1997, \aap, 327, 901 
\bibitem[risa05]{} Risaliti, G., et al. 2005, MNRAS, accepted (R05, astro-ph/0510861)
\bibitem[Risaliti \& Elvis(2005)]{2005ApJ...629L..17R} Risaliti, G., \& 
Elvis, M.\ 2005, \apjl, 629, L17 
\bibitem[Spergel et al.(2003)]{2003ApJS..148..175S} Spergel, D.~N., et al.\
2003, \apjs, 148, 175
\bibitem[Spoon et al.(2003)]{2003A&A...402..499S} Spoon, H.~W.~W., 
Moorwood, A.~F.~M., Pontoppidan, K.~M., Cami, J., Kregel, M., Lutz, D., \& 
Tielens, A.~G.~G.~M.\ 2003, \aap, 402, 499 
\bibitem[Vignati et al.(1999)]{1999A&A...349L..57V} Vignati, P., et al.\ 
1999, \aap, 349, L57 
\end{thebibliography}
\end{document}